# A conjecture of thermo-gravitation based on geometry, classical physics and classical thermodynamics

**The authors:** Weicong Xu [a, b], Li Zhao [a, b, *]

[a] Key Laboratory of Efficient Utilization of Low and Medium Grade Energy, Ministry of Education of China, Tianjin 300350, China

[b] School of mechanical engineering, Tianjin University, Tianjin 300350, China

* Corresponding author. Tel: +86-022-27890051; Fax: +86-022-27404188; E-mail: jons@tju.edu.cn

## Abstract

One of the goals that physicists have been pursuing is to get the same explanation from different angles for the same phenomenon, so as to realize the unity of basic physical laws. Geometry, classical mechanics and classical thermodynamics are three relatively old disciplines. Their research methods and perspectives for the same phenomenon are quite different. However, there must be some undetermined connections and symmetries among them. In previous studies, there is a lack of horizontal analogical research on the basic theories of different disciplines, but revealing the deep connections between them will help to deepen the understanding of the existing system and promote the common development of multiple disciplines. Using the method of analogy analysis, five basic axioms of geometry, four laws of classical mechanics and four laws of thermodynamics are compared and analyzed. The similarity and relevance of basic laws between different disciplines is proposed. Then, by comparing the axiom of circle in geometry and Newton's law of universal gravitation, the conjecture of the law of thermo-gravitation is put forward.

## 1. Introduction

With the development of science, the theories and methods of describing the same macro system are becoming more and more abundant. It seems that a set of unified and mutually confirmed scientific theories is the unified pursuit of human cognition of the world. There are two ways to follow in the development of modern physics. One is to find new phenomena through experimental research, then summarize the law through a large number of experimental data, and finally develop the physical theory. The other is to predict the possible physical phenomena through theoretical derivation, and finally verify the physical theory through experimental research.

As for the most basic physical theory, it is difficult to carry out experimental research. It is an important way to develop the basic theory of physics by analogy with other more mature disciplines.

Geometry has experienced the development of nearly 30 centuries, and the basic knowledge of modern geometry still effectively guides the development of human society, which proves the correctness of geometry. In the 7th century BC, geometry was born in ancient Egypt [1]. Later, it went through the development of ancient Greek scholars and Pythagoras. However, during this period, the knowledge of geometry is relatively scattered, and a relatively complete theoretical system has not yet been formed. It was not until 300 BC that Euclid published the 'Euclid's Elements' [2], which was the first time to realize a more systematic exposition of geometry knowledge. According to the 'Euclid's Elements', the theoretical system of geometry is based on five axiomatic assumptions, and thousands of theorems and laws are deduced, which have been guiding the development of society.

In the field of physics, classical mechanics has been fully developed, which is an effective method to study and describe the motion law of objects in the state of macroscopic world, far less than the speed of light and weak gravitational field before relativity and quantum mechanics [3]. The basic law of classical mechanics is Newton's law or other equivalent mechanical principles. In 1678, the 'Philosophiae Naturalis Principia Mathematica' was published by Newton [4], which is the basic of classical mechanics. In this book, Newton's first law, second law, third law and Newton's law of universal gravitation were proposed for the first time. Classical mechanics is not only the basis of classical physics and astronomy, but also the theoretical basis of modern engineering mechanics and related engineering technology.

Classical thermodynamics is based on the four laws of thermodynamics, using thermodynamic data to study the relationship between the macro properties of the equilibrium system and reveal the direction and limit of processes. It does not involve the microscopic properties of particles. The establishment of classical thermodynamics is a long and arduous process [5]. Many physicists have made their own indispensable contributions. Among them, the theory of heat has undergone a tortuous historical evolution of mutual replacement. In 1842, J. R. von Mayer put forward the theory of conservation of energy, which recognized that heat is a form of energy, which can be transformed into mechanical energy, and calculated the thermal work equivalent from the difference between the specific heat capacity at constant pressure and that at constant volume. J.P. Joule, a British physicist, established the concept of electrothermal equivalent in 1840. After 1842, he measured the thermal equivalent in different ways. In 1850, Joule's experimental results made the scientific community completely abandon the theory of heat and mass, and recognized that the first law of thermodynamics, which conserves energy and can be exchanged in the form of energy, is an objective natural law. In 1850 and 1851, Clausius and Kelvin from Germany put forward the second law of thermodynamics, and on this basis, they proved Carnot's theorem again. Clausius gave the definition of entropy and put forward the principle of entropy increase of isolated system. The third law of thermodynamics is usually expressed as absolute zero, the entropy of all perfect crystals of pure matter is zero, or absolute zero is not attainable. The four basic laws of thermodynamics are the basis of the macro theory of thermal phenomena, and play an indelible role in the development of industrial production and even society.

To sum up, the research objects of geometry, classical mechanics and classical thermodynamics are all macro systems. Although the research methods are different, with the gradual completion of the theoretical system, the basic postulates and laws supporting different disciplines are bound to have relevance or even similarity. Among these three disciplines, geometry has the longest history, and its theoretical system should be the most complete. Although two different branches were developed [6], the original theory always maintains a high degree of self-consistency. Therefore, it is possible to use geometry to guide the theoretical basis of classical mechanics and thermodynamics.

2. Method

2.1 Axioms of geometry

Axiom 1: Between any two points, a straight line may be drawn.

Axiom 2: Any terminated straight line can be infinitely extended from both ends into a straight line.

Axiom 3: A circle may be drawn with any point as its center and any distance as its radius.

Axiom 4: All right angles are equal.

Axiom 5: if a line intersects two lines and the sum of two internal angles intersected on the same side is less than two right angles, then the two lines must intersect at a point on that side after infinite extension.

The fifth axiom can be equivalent to: through a point outside the line, only one line can be drawn parallel to the original line. In 1826, Lobachevsky put forward that through a point outside the line, there can be countless lines parallel to the original line, which started the development of Lobachevsky geometry [7]. In 1851, Riemannian presented that through a point outside the line, no line is parallel to the original line, which opens up Riemannian geometry [8].

**2.2 Laws of classical mechanics**

Newton's first law: If an object does not interact with other objects, it is possible to identify a reference frame in which the object has zero acceleration.

Newton's second law: When viewed from an inertial reference frame, the acceleration of an object is directly proportional to the net force acting on it and inversely proportional to its mass.

Newton's third law: If two objects interact, the force $F_{12}$ exerted by object 1 on object 2 is equal in magnitude and opposite in direction to the force $F_{21}$ exerted by object 2 on object 1.

Newton's law of universal gravitation: Every particle in the universe attracts every other particle with a force that is directly proportional to the product of their masses and inversely proportional to the square of the distance between them.

**2.3 Laws of classical thermodynamics**

The first law of thermodynamics: There are various forms of energy, which can be transformed from one form to another. In the process of transformation, the amount of energy remains unchanged.

The second law of thermodynamics: It is impossible to transfer heat from a low-temperature object to a high-temperature object without causing other changes; It is impossible to take heat from a single heat source and make it completely change into useful work without

other effects

The third law of thermodynamics: It is impossible to make the temperature of an object equal to absolute zero.

The zeroth law of Thermodynamics: If each of the two thermodynamic systems is in thermal equilibrium (at the same temperature) with the third thermodynamic system, they must also be in thermal equilibrium with each other.

**2.4 The relationship among axioms of geometry, laws of classical mechanics and laws of thermodynamics**

Through comparison, it is found that there are correlations among geometry axioms, laws of classical mechanics and laws of thermodynamics. The specific analysis is as follows:

The first correlation exists between axiom 1 of geometry, Newton's second law and the second law of thermodynamics. Based on the axiom 1 of geometry, a straight line can be drawn through two points. The equation of a straight line can be expressed as $y = kx$, where the slope $k$ can be determined by two points. This axiom 1 of geometry can be understood as that the dependent variable is proportional to the change of the independent variable. As for the Newton's second law, the equation can be expressed as $F=ma$, where $F$ represents the force exerted on the object, $m$ represents the mass of the object, and $a$ represents the acceleration of the object. Newton's second law can be understood as the force is proportional to the acceleration of the object. The second law of thermodynamics expresses the proportional relationship between the heat absorbed by the thermal system and the entropy change, and its equation can be expressed as $\delta Q = T dS$, where $Q$ represents the absorbed heat, $s$ represents the entropy, and $T$ represents the temperature of the thermal system. To sum up, the axiom 1 of geometry, Newton's second law and the second law of thermodynamics all represent the proportional relationship between two physical quantities.

The second correlation exists between axiom 2 of geometry, Newton's first law and the first law of thermodynamics. They all indicate that a certain physical quantity remains unchanged. The axiom 2 of geometry clarifies that a line segment can be extended to a straight line, that is, the slope of the line segment remains unchanged. Newton's first law indicate that the motion state of the object remains unchanged if the object is not subjected to external force. In the process of energy conversion, the total amount of energy remains unchanged, which is stated in the first law of thermodynamics.

The third correlation exists between axiom 4 of geometry, Newton's third law and the zeroth law of thermodynamics. The axiom 4 of geometry means that all right angles are equal. It can be understood that angle A equals angle B, angle B equals angle C, then angle A equals angle C. The expression of right angle could be considered as a special case. Newton's third law is expressed as $\vec{F_{12}} = -\vec{F_{21}}$, where $\vec{F_{12}}$ represents the force exerted by particle 1 on particle 2, $\vec{F_{21}}$ represents the reaction force of particle 2 on particle 1. This law describes the relationship between the forces of two particles, and can also be extended to the forces of more than three particles, that is, $\vec{F_{12}} = -\vec{F_{21}}$, $\vec{F_{23}} = -\vec{F_{32}}$, then $\vec{F_{12}} = -\vec{F_{32}}$. Although the force is a vector and has a directional representation, it is essentially a balanced transfer of force. As for the zeroth law of thermodynamics, if the temperatures of system A and system B are equal to that of system C, then the temperatures of system A and system B are also equal. The axiom 4 of geometry, Newton's third law and the zeroth law of thermodynamics represent the properties of equilibrium transfer between physical quantities

## 3. Results and discussion

### 3.1 The conjecture of thermo-gravitation

Through the above analysis, it is obvious that there is no analogy between the third law of thermodynamics in classical thermodynamics, the law of gravitation in classical mechanics and axiom 4 and 5 of geometry. The next part of this paper will introduce the new law of thermodynamics by comparing the axiom 3 of geometry and Newton's law of universal gravitation.

The expression of gravitation is $F = G\dfrac{m_1 m_2}{r^2}$, where $F$ represents the gravitational force between two objects, $m_1$ and $m_2$ represent the mass of two objects respectively, and $r$ represents the distance between two objects. If $m_1 = m_2$, then $F = G\left(\dfrac{m}{r}\right)^2$, which is similar to the expression of the area of the circle, that is $s = \pi r^2$, where s represents the area of a circle, r represents the radius of the circle. It is speculated that the expression of the new law of thermodynamics corresponding to the axiom 3 of geometry and Newton's law of universal gravitation should also be similar to $y=Cx^2$.

In addition, according to Einstein's basic theory of physics, energy and mass have the same essence, so we can imagine that energy can reflect the nature of mass. There is universal gravitation between objects with non-zero mass and other objects with mass in the universe. Heat is a form of energy. Will it also have the characteristics of gravitation?

The temperature of a thermodynamic system is a statistical result, which is related to the average kinetic energy of the molecules involved in the thermodynamic system. The kinetic energy is a function of velocity. Therefore, the temperature of the thermal system is a function of velocity. In addition, the entropy of the thermodynamic system increases unidirectionally in the process of spontaneous change, which is consistent with the characteristics of time. Therefore, there must be some functional relationship between entropy and time.

Based on the Albert Einstein's special-relativity equation $\Delta E=\Delta mc^2$ and Newton's second law $\Delta F=\Delta ma$, the expression of the new law of thermodynamics, which is named as the law of thermo-gravitation, can be expressed as:

$$\Delta F=\Delta ma=\frac{\Delta E}{c^2}\frac{\Delta v}{\Delta t} \qquad (1)$$

where $v$ represents the velocity and $t$ represents the time. As mentioned above, velocity is a function of temperature and time is a function of entropy. At the same time, assuming that all energy changes are heat energy, formula (1) can be deduced as:

$$\Delta F=\frac{\Delta E}{c^2}\frac{\Delta v}{\Delta t}=\frac{T\Delta S}{c^2}\frac{f(\Delta T)}{f(\Delta S)}=Z\left(\frac{T}{c}\right)^2 \qquad (2)$$

where Z represents a constant of thermo-gravitation.

**3.2 Possible application scenarios of thermo-gravitation**

Gravitation is the universal force between any two objects in space, which guides the development of society and explains the motion of planets in space. However, the accurate measurement of gravitational constant G is always a complicated problem in physics, which is a special key parameter in branches of theoretical physics, astrophysics and geophysics. At present, the measurement accuracy of gravitational constant G is the lowest among the basic constants of physics. According to the literature report in 2018[9], the highest measurement accuracy of G value could reach $10^{-5}$ level. Same as universal gravitation, thermo-gravitation exists between any two objects in space. A further conjecture is that the force between any two objects in space should be the sum of universal gravitation and thermo-gravitation. This possible lead to the inaccurate measurement result of G value at this stage.

How long the sun can exist is a problem that people have been concerned about, and it is also a frontier problem in physics. At this stage, the accepted result is that the sun still has 5 billion years to live. In the calculation of the existence time of the sun, the mass of the sun is the first parameter to be determined, which is calculated by the gravitational force between the earth and the sun. However, as mentioned above, there should be thermo-gravitation between any two objects in space. The temperature difference between the sun and the earth is more than 5000K, therefore, the thermo-gravitation cannot be ignored, which directly determines the mass of the sun. According to the above conjecture, the gravitational force between the sun and the earth should be less than the current measured value. Therefore, the mass of the sun should be smaller than the current value. As a result, the sun probably will exist for less than 500 million years.

## 4. Conclusion

In this paper, by analogy with the core laws of geometry, classical mechanics and classical thermodynamics, the conjecture of new laws of thermodynamics, that is the law of thermo-gravitation, is put forward. The specific conclusions are as follows:

(1) Through comparative analysis, it is found that the axiom of straight line in geometry, Newton's second law and the second law of thermodynamics are similar, which indicate the proportional relationship between the two variables. The axiom of line segment in geometry, Newton's first law and the first law of thermodynamics are similar, which reflect the constancy of a parameter. The axiom of angle in geometry, Newton's third law and the zeroth law of thermodynamics are similar, which reflects the equivalence between parameters.

(2) By comparing and analyzing the axiom of circle in geometry and Newton's law of universal gravitation, the law of thermo-gravitation in thermodynamics is put forward, which is expressed as $\Delta F = Z \left( \dfrac{T}{c} \right)^2$. What's more, the physical problems related to thermo-gravitation are proposed.